\documentclass[useAMS]{mn2e}

\usepackage{graphicx}	
\usepackage{amsmath}	
\usepackage{amssymb}	
\usepackage{multicol}       
\usepackage{bm}		
\usepackage{pdflscape}	
\usepackage[utf8]{inputenc}
\usepackage[T1]{fontenc}
\usepackage{ae,aecompl}

\title[Review of recent wide binary gravity tests] {A critical review of recent GAIA wide binary gravity tests} 

\author[X. Hernandez, Kyu-Hyun Chae and A. Aguayo-Ortiz] {X. Hernandez$^{1}$\thanks{xavier@astro.unam.mx}, Kyu-Hyun Chae$^{2}$\thanks{chae@sejong.ac.kr}
 and A. Aguayo-Ortiz$^{1}$ \\ 
$^{1}$ Instituto de Astronom\'{\i}a, Universidad Nacional Aut\'{o}noma de M\'{e}xico,
Apartado Postal 70--264 C.P. 04510 M\'exico D.F. M\'exico.\\
$^{2}$ Department of Physics and Astronomy, Sejong University, 209 Neungdong-ro Gwangjin-gu, Seoul 05006, Republic of Korea.\\
}

\date{Released 10/12/2023}

\begin{document}

\label{firstpage}

\maketitle

\begin{abstract}

  Over the last couple of years, the appearance of the third data release from the {\it Gaia} satellite has triggered various wide binary low acceleration
  gravity tests. Wide binaries with typical total masses $\approx 1.0 - 1.6 M_{\odot}$ and separations above a few thousand au probe the low acceleration
  $a \lesssim a_{0}$ regime, where at galactic and larger scales gravitational anomalies typically attributed to the presence of an as yet undetected dark
  matter component appear, where $a_{0} \approx 1.2\times 10^{-10}$ m s$^{-2}$ is the acceleration scale of MOND. Thus, studies of the relative velocities and
  separations on the plane of the sky, $v_{2D}$ and $s_{2D}$ respectively, of wide binary stars extending to separations above a few kau, provide an independent
  approach on the empirical study of gravity in the interesting  $a \lesssim a_{0}$ acceleration range. Two independent groups, through complementary approaches,
  have obtained evidence for a departure from Newtonian predictions in the low acceleration regime, in consistency with MOND expectations for wide binary orbits
  in the Solar Neighbourhood. Two other groups however, have instead reported results showing a clear preference for Newtonian gravity over various MOND
  alternatives tested, over the same low acceleration regime. We here take a critical look at the various studies in question, from sample selection to statistical
  treatment of the wide binary relative velocities obtained. We discover a couple of critical problems in the formal design and statistical implementation
  shared by the two latter groups, and show explicitly how these yield biased conclusions.
\end{abstract}


\section{Introduction}

One of the most relevant scientific debates of the last few decades has been the discussion of whether to attribute the well established
gravitational anomalies appearing in the low acceleration regime of $a<a_{0}$ (where $a_{0} \approx 1.2 \times 10^{-10}$ m s$^{-2}$) at galactic and
cosmological scales to the presence of an undetected dark matter component, or { as} evidence for a change in the behaviour of gravity at such
low acceleration scales not probed by any direct experiment. In an attempt to obtain independent evidence on the low acceleration behaviour
of gravity at scales where dark matter is not envisioned to play any role, one of us proposed in Hernandez et al. (2012) to directly test
gravity using statistical observations of large samples of Solar Neighbourhood wide binary stars. For solar mass wide binary stars, the low
acceleration regime where anomalies are apparent in the galactic and cosmological context appears above separations of a few kau. { Fortunately,
there are many such systems. Given the quality of the proper motion data available at the time, those first tests proved inconclusive.}

With the appearance of the {\it Gaia} EDR3 and DR3 catalogues the situation changed, and parallax, proper motion, photometric and spectroscopic
data of sufficient quality to allow the inference of relative velocities on the plane of the sky, $v_{2D}$, separation on the plane of the sky,
$s_{2D}$, as well as high quality stellar masses, $M_{\star}$, inferred from stellar evolutionary models, and in many cases even radial velocities,
became available for large samples of local wide binary stars with $s_{2D}$ values reaching up to several tens of kau. Once data of the required
quality { became} available, one still { had} to deal with the statistical nature of the problem, given the long orbital periods of the systems in question,
which make the actual tracing of individual wide binary orbits impractical given the current accuracy of observations. Statistical samples
of $v_{2D}$, $s_{2D}$ and $M_{\star}$ can then in principle be used to decide if gravity is behaving in a constant manner, { or not}, across the range of $s_{2D}$
values sampled.

However, deciding if there is any change, or not, in the law of gravity from a statistical sample of $v_{2D}, s_{2D}$ and total stellar masses, $M_{T}$,
still requires { handling} the presence of kinematic contaminants and possible external perturbers. A given wide binary star might harbour in one
of its members (or both-in the interest of brevity the distinction will not further be made explicit if not needed) a hidden tertiary component. { This}
might not be apparent from current observations, if this extra component is dim enough and close enough to one of the readily identified { members} of
the wide binary. The presence of such hidden tertiaries leads to an error in the assigned total mass of the wide binary in question, and also, due to
the motion of the internal binary, produces a kinematic offset on the observed wide binary, with respect to the motion that one is trying to measure.
In some cases, an apparent wide binary star might simply be the result of projection effects in a crowded field, or a flyby between unbound field stars.
Then, in the case of loosely bound very wide pairs, it becomes increasingly difficult with increasing $s_{2D}$ values, to find isolated systems which
can be guaranteed not to show perturbed kinematics as a result of interactions with nearby field stars. Further, the details of the statistical implementation
used to gauge the presence or otherwise of a deviation from Newtonian expectations is not uniquely established, and the details of it might bias the
results obtained, if one is not careful to test the full scheme being used and if no consistency safeguards are introduced in the approach.

All of the above points result in study design details which have been chosen differently by various groups. Here we critically
review these varied choices from all the groups currently working in the field. We examine sample selection strategies, result presentation choices and
statistical inference methods in detail, to understand the divergence of published results, and suggest improvements and consistency checks useful
towards reaching a definitive answer in the field.

In Section 2 we describe the recent {\it Gaia} wide binary tests yielding consistent results in favour of MOND for the low acceleration regime, while
Sections 3 and 4 discuss the implementation of tests resulting in a reported preference for Newtonian gravity over MOND options in this same acceleration regime.
We identify two main serious shortcomings shared by the latter groups, which we explore in Section 5 to demonstrate explicitly how and why these lead to
biased results, which the lack of consistency checks failed to identify. Finally, Section 6 presents our conclusions.

\section{Recent results showing clear MOND low acceleration phenomenology}

The first part of the wide binary gravity test in all cases is the sample selection process. In all the studies reviewed here, the initial steps of
this process are taken from the ideas of El-Badry \& Rix (2018) and El-Badry et al. (2021). Making use of the {\it Gaia} catalogues, a preliminary list of
binary candidates is constructed from all pairs within a { certain distance from the Sun, chosen to satisfy} that the distance between stars in the pair along the line
of sight lies { (within errors)} within a factor of order 2 of the separation on the plane of the sky of the pair. Then, all potential pairs containing individual
stars which are members of more than one pair are excluded. { These points are} common to all the samples discussed below. { Next}, binaries having nearby perturbers
are excluded, the details of this step change from one study to another, varying on how strict an exclusion this step becomes, as discussed below. The next important
difference is in the range of separations on the plane of the sky between stars in the binaries to be considered by the various authors, this point will also be explored
in detail. Finally, a series of quality cuts and exclusion of kinematic contaminants are introduced, again, to varying degrees of strictness, as detailed below.

\subsection {Work by the Hernandez et al. group}

We begin with a description of the results presented by the group headed by one of us, Hernandez and collaborators. In Hernandez et al. (2019) using the first
{\it Gaia} DR2 catalogue this group identified strong indications of a regime change in $v_{1D}$ vs. $s_{2D}$ for a small sample of 81 wide binaries ranging in
projected separations from 0.03-1 pc, 0.6-200 kau. This wide binaries had been selected from the Shaya \& Olling (2011) carefully selected sample of wide binaries
from the Hipparcos database originally used by Hernandez et al. (2012), and were revisited in Hernandez et al. (2019) updating the astrometry using the {\it Gaia} DR2
catalogue. Although the strict selection cuts eliminating groups and projection effects in the Shaya \& Olling (2011) catalogue, plus the large fraction of Hipparcos
stars missing from the {\it Gaia} DR2 catalogue resulted in only a small sample with poor statistics, the closest binaries appeared consistent with the detailed
Newtonian predictions of Jiang \& Tremaine (2010). This last reference calculated detailed expectations for the relative velocity distribution of wide binaries
in the Solar Neighbourhood under Newtonian expectations.
The widest binaries however, appeared to show a degree of enhanced relative velocity, despite the large error bars. The difficulty of finding
isolated wide binaries towards the widest range in Hernandez et al. (2019), given the average inter-stellar separation in the Solar Neighbourhood is of
1pc, and the need to keep clear off the tidal radius for such systems, of $<M_{T}> \approx 1.6 M_{\odot}$, also of order 1pc, suggested reducing substantially the
upper range for future studies. On the other hand, it appeared convenient to include a deep Newtonian regime where even MOND expectations coincided with Newtonian
predictions, so as to have an internal consistency check for all the astrometry and statistical procedures of the test. For these reasons in Hernandez et al. (2022)
(henceforth H22) it was decided to focus on the 0.2-12 kau separation range, using {\it Gaia} EDR3.

For the above reasons in H22, and all subsequent papers by that group, the wide binary projected separation range of study was fixed to
0.2-12 kau. Following the philosophy of keeping a maximum control on all the systematics of the problem, at the expense of limited statistics, extremely strict
hidden tertiary exclusion criteria were followed. Then, a strict isolation criterion was applied to exclude any binary
candidate in the vicinity of any other {\it Gaia} source, out to 0.5 pc. Thus, even for the widest binaries considered, at 12 kau, 0.06 pc, the nearest {\it Gaia} neighbour
must lie more than 8.3 times the binary internal separation away. Data quality cuts follow to ensure only binaries where both stars have a RUWE internal {\it Gaia}
single star solution quality index $<1.2$, imposing a final distance cut of 130 pc. All stars included in the final sample are main sequence stars and have {\it Gaia}
reported radial velocities, which ensures high quality spectroscopic single star solutions, and therefore a low probability of harbouring hidden tertiaries.
A strict HR diagram cut follows, to guarantee that only stars in regions of the HR diagram { where} the probability of containing hidden tertiaries is minimal, as per
independent assessments of this in Belokurov et al. (2020) and Penoyre et al. (2020). High signal-to-noise ratios in parallax and proper motions are then required,
to further cut the sample. Finally, relative velocity filters both on the plane of the sky and along the line of sight of $<4$km s$^{-1}$ are introduced, to exclude
flybys and projection effects, as the two point relative velocity distribution for pairs of field stars in the Solar Neighbourhood is a Gaussian distribution with
a dispersion of $>$ 60 km s$^{-1}$, for the old main sequence stars being studied. The final sample contained 423 binary pairs with average signal-to-noise ratios
{ of} internal relative velocities of 14.88 and 18.62 for RA and Dec, respectively and mean $M_{T}=1.6 M_{\odot}$. In all other studies by this same group selection
criteria are minor variations of those described above, resulting in small variations in final total numbers and internal relative velocity errors. In describing
those other studies only the final binary numbers and signal-to-noise ratios in internal relative velocities will be given, details can be found in the papers
in question.

In H22 results are presented as scatter plots of $v_{1D}$ vs. $s_{2D}$, including also binned rms values of these quantities, for RA and DEC observations treated
separately. { These} binned values are compared to the same quantities as estimated for a Newtonian galactic model by Jiang \& Tremaine (2010). The comparison
shows accordance between the inferred data and the Newtonian predictions for small separations $<2$kau, without any parameter fitting. However, data deviate
upwards of the Newtonian prediction for larger separations, with rms values remaining at a constant level of 0.45 km s$^{-1}$. No quantitative analysis of these
results were presented in H22, with the aim having been only to establish or reject the presence of a gravitational anomaly in the low acceleration regime.
In H22 the anomaly seen in the rms binned $v_{1D}$ values was mistakenly interpreted as evidence for deep MOND behaviour, analogous to the flat rotation curve
regime in galaxies. 

Next, using the full DR3 {\it Gaia} catalogue, Hernandez (2023) repeated the experiment using 450 binary stars in the 0.2-12 kau separation range, all main sequence
stars within 125 pc from the Sun, with final mean signal-to-noise ratios in velocity of 15.7. The results accurately traced what was reported in H22, but this
time the presentation included mean and median $v_{2D}$ values as well as rms $v_{1D}$ ones. The two former were consistent with detailed MOND expectations for a
change in the effective value of the gravitational constant of $G\to \gamma G$, with $\gamma \approx 1.4$. No quantitative assessment of this model was attempted
and only qualitative results were presented, showing however a clear deviation from Newtonian expectations in the $s_{2D}>2000$ au low acceleration regime,
see the left panel of Fig.4, Fig. 8 and the right panel of Fig. A1 in that paper.

Lastly, Hernandez et al. (2024) -henceforth H24- used 574 binary stars in the 0.2-12 kau separation range, all main sequence stars within 125 pc from the Sun,
with final mean signal-to-noise ratios in velocity of 16.4. This time to perform a detailed statistical study of the $v_{2D}$ vs. $s_{2D}$ distribution and compare to
$G \to \gamma G$ models, to obtain best fit $\gamma$ values as a function of acceleration range. 
In H24 a careful statistical analysis to find optimal values of the effective gravitational constant, $G \to \gamma G$,
relevant to a clean sample of observed wide binaries was developed. This study works using the dimensionless velocity variable $\tilde{v}=(G M_{T}/s_{2D})^{-1/2}v_{2D}$,
and takes advantage of the extremely high quality $s_{2D}$ data and highly accurate stellar mass estimates available from using {\it Gaia} data for nearby
systems, typically within about 100 pc, so that the error budget on $\tilde{v}$ is overwhelmingly dominated by uncertainties in $v_{2D}$. The method developed
in this paper fully accounts for all projection effects of intrinsic wide binary orbits onto the plane of the sky, taken as probability density functions
under the assumption of isotropy, as well as the probability density function of orbital phase occupancy and the probability density function of ellipticity
distributions, working with the dimensionless 2D relative velocity $\tilde{v}$ mentioned above. Results were calculated for a range of statistical ellipticity
distributions as inferred directly by Hwang et al. (2022) for Solar Neighbourhood {\it Gaia} wide binaries, and the resulting variations on the final inferences
included as a systematic contribution to the final error budget quoted. This last reference parameterised the probability distribution of ellipticities for Solar
Neighbourhood wide binaries as $ P(e)=(1+\alpha) e^{\alpha}$, where the dimensionless parameter $\alpha$ was determined by those same authors to lie in the range
$1.0<\alpha<1.4$ for local {\it Gaia} wide binaries. H24 then obtained full accordance with Newtonian expectation in the high acceleration $s_{2D}<2$ kau
region, as a direct result of the method used, in the total absence of any parameter fitting, yielding $\gamma=1.00 \pm 0.1$. Again, for the low acceleration
$s_{2D}>2000$ au region, a clear gravitational anomaly is evident, results are consistent with MOND expectations of $\gamma=1.5 \pm 0.2$.

It is important to realise that the data to be used are not error-free. Indeed, we will always be working with relative velocity distributions which necessarily include a
level of observational noise, and hence, which are broadened versions of an intrinsic reality. This broadening, while not exactly symmetric to the left and right of
intrinsic $v_{2D}$ and $\tilde{v}$ distributions, which are not exactly symmetric, will apply almost equally towards smaller and larger values with respect to intrinsic
velocity distributions. A further effect of noise will be to introduce a variance, a fixed input model will not be modified in a unique deterministic manner, but only
in a probabilistic one. This is relevant when considering small samples, such as the ones in H24 which contain only 108 binaries in the low acceleration region, or
when larger samples are binned into many tens or even hundreds of smaller cells, each of which will end up containing only a small number of observational points.
For these reasons, H24 included extensive testing with synthetic samples to ensure that no biases enter the procedure. The confidence interval quoted above includes all
statistical, systematic and resolution uncertainties entering the problem, as resulting from careful error propagation analysis and repeated Monte Carlo (MC)
resampling of the data within their {\it Gaia} reported uncertainties, to fully account for the variance present in the test performed.

\subsection{Work by K-H. Chae}

We now turn to the independent studies of Chae (2023), Chae (2024a) and Chae (2024b), henceforth C23, C24a and C24b.
C23 presented results for a large sample of 26,615 {\it Gaia} DR3 wide binaries within 200 pc from the Sun in the range
of $200<(s_{2D}/au)<30,000$, selected through a careful isolation criterion to ensure the absence of flyby events and external perturbations, to test
for changes in gravity across the separation range probed. A careful statistical de-projection method is implemented such that each observed
($s_{2D}$, $v_{2D}$) wide binary data point is turned into { a} statistical distribution of possible ($s_{3D}$, $v_{3D}$) values by taking into account isotropic projection
probabilities, the time and velocity dependence of the orbital dynamics in question to obtain orbital phase occupancy probability distributions, and individual
ellipticity estimates for the binaries used, derived from the observed angles between $s_{2D}$ and $v_{2D}$ for these same binaries, from Hwang et al. (2022).
{ Hwang} et al. (2022) estimate ellipticities for local wide binaries from the {\it Gaia} satelite through the angle between the projected relatve possitions and the
projected relative velocities on the plane of the sky, albeit assuming no hidden tertiaries are present. These last ellipticities are treated as probability density
functions, given their reported confidence intervals.

This sample is cleared of resolved triples and higher-order multiples, but
not cleared of hidden tertiaries or hidden higher-order multiples, and hence their effect must be modelled before drawing any inferences on the behaviour of gravity.
The details of { how} incuding hidden tertiaries modify $v_{2D}$ and $\tilde{v}$ distributions are important and must be appreciated to fully understand
the problem. Firstly, the addition of an unacknowledged extra stellar component implies the addition of extra mass, hence $v_{2D}$ and $\tilde{v}$ distributions
will be shifted to larger values in samples containing hidden tertiaries, with respect to samples clean of these contaminants, at a fixed underlying gravity
law. Then, the kinematic effect of the inner binary, when added as a vector sum to the relative motion of the wide binary, will result in a correction which
can be positive or negative. In going to the wide binaries, where relative velocities become smaller with growing separations, the kinematic contaminant
of the inner binary will generally dominate, and hence this effect too will generally result in a shift to larger relative velocity values. This is illustrated
clearly in Fig. 10 of { Banik et al. (2024) (henceforth B24)}, where $\tilde{v}$ distributions are given for a MOND model (blue curves) and for a MOND model after
the addition of 65.7 \% of hidden tertiaries to the sample (red curves), for a set of four assumed $s_{2D}$ bins. It is evident that the addition of hidden tertiaries
biases the resulting distribution, resulting in a broadening towards higher velocity values, and an overall shift towards higher values, the lowest velocity
values in the original distribution are transformed into larger values, as happens to the larger values as well. Hence, after adding hidden tertiaries,
the lower velocity ranges of the distribution are de-populated, { resulting in} an overall broadening towards larger values.

In C23 the modelling of hidden tertiaries is achieved { using statistical models of wide binaries, where to a fraction $f_{\rm{multi}}$ of which} an extra close
component is added (a triple to quadruple ratio of 7:3 was used), and after a probabilistic projection of all orbital elements is performed, the change in mass and $v_{2D}$
distributions is calculated. The resulting $v_{2D}$ distributions are then compared against the observed sample in the high acceleration region, where Newtonian
dynamics applies, to calibrate $f_{\rm{multi}}$. Thus, the only kinematic contaminant remaining in the sample is firmly calibrated through comparison with Newtonian
expectations. Since resolved multiples have been removed, the only multiple systems of concern remaining are unresolved ones. The validity of the scheme implemented
is hence assured by the statistical independence of the probability of hosting an unresolved companion with $s_{2D}$, provided that stars' masses are statistically
independent of $s_{2D}$ and all stars are selected with the same criteria. { These conditions hold in C23.} Indeed, in a study of 4947 {\it Gaia} wide binaries by
Hartmann et al. (2022) it was found that $f_{\rm{multi}}$ for all tertiaries (and quarternaries) including resolved ones is very weakly dependent on $s_{2D}$. See also
Tokovinin et al. (2002), Tokovinin et al. (2010), and Tokovinin (2014) for results from nearby surveys.
Once this { single} free parameter has been calibrated in the high acceleration Newtonian region, results for the low acceleration
$s_{2D}>2000$ au region clearly show a break in the behaviour of gravity of the type $G \to \gamma G$ with $\gamma =1.43 \pm 0.06$, e.g. Fig. 19 in C23,
ruling out a low acceleration Newtonian behaviour and consistent with MOND AQUAL expectations of $\gamma \approx 1.4$. A high quality sub-sample
with a distance limit of just 80 pc is also included as a check, and it yielded consistent results. Given the close similarity of expectations under
AQUAL and QMOND models for the problem being treated (Banik \& Zhao 2018), henceforth any mention to AQUAL has to be understood as applying also to
QMOND MOND variants.

C24a further considered a much smaller sample (about 10 percent of the C23 sample) of binaries selected with much stricter astrometric and kinematic
criteria to include only ``pure'' binaries statistically free of hidden close companions. Because at least 50 percent of the C23 sample are expected to be free of
hidden close companions, picking only 10 percent represents a conservative safe choice. This is indeed shown to be the case through an accurate fit to Newtonian
expectations across the $200<(s_{2D}/au)<1000$ range, without the need of fitting any free parameters or of including any hidden components, i.e., using
$f_{\rm{multi}}=0.0$. For this restrictive sample, C24a tested gravity through the observed scaling of $v_{2D}$ with $s_{2D}$ as well as the MC
reconstructed 3D acceleration defined in C23.


This new sample confirmed the C23 results with $\gamma=1.49 \pm 0.2$ (and a velocity boost factor of $\gamma_v = 1.20 \pm 0.06$ satisfying the expected
relation $\gamma = \gamma_v^2$) for the low acceleration $s_{2D}>5000$ au regime, again ruling out Newtonian dynamics in this regime, and fully consistent with AQUAL
MOND expectations, e.g. Fig. 11 in that paper. This paper includes a small correction to the results of C23 as a consequence of including a correction to a small bug
in the code developed in C23 (see Appendix A in C24a), which lead to an artificially high $f_{\rm{multi}}$ value reported in C23. This crucial factor is estimated to be $0.2
\lesssim f_{\rm{multi}} \lesssim 0.5$ for various sub-samples using the corrected code. This result is broadly consistent with estimates from near-by surveys
(see reviews by, e.g.,  Offner et al. (2022), Moe \& Di Stefano (2017)) that typically find $f_{\rm{multi}}<0.5$ including even resolved multiples. Contrary
to B24's criticism of the results in C23, we note that the points on the acceleration plane from MC de-projections of each observed wide binary are probability
distributions taking into account all possible observational ranges of inclinations, eccentricities, hidden close companions, etc. Similar MC approaches
are used in the analyses of velocities by C24a and H24. Hence, criticisms found in B24 based exclusively on comparisons of the median $\tilde{v}$ parameter are not applicable
to studies considering the full distribution of modelled and observed velocities or accelerations. Notice also that in C24a various progressive data quality
cuts are tested, all yielding consistent results, showing that the study is robust with respect to the level of observational noise included. { The} removal of
the lower quality data sub-samples has no impact on the conclusions.

 \begin{figure*}
 \vskip 0pt
 \hskip -5pt
 \includegraphics[height=7.0cm,width=8.7cm]{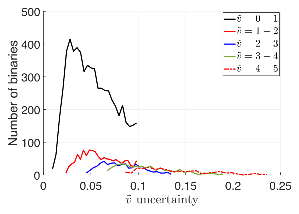}
 \hskip 5pt \includegraphics[height=7.0cm,width=8.7cm]{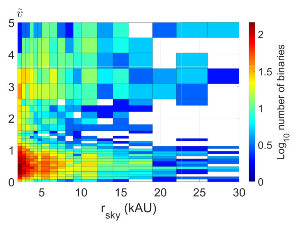}
 \caption{Left: Distribution of errors in $\tilde{v}$ as a function of the ranges for the values of this parameter in Banik et al. (2024). Notice that for the
   crucial first two ranges, errors of above 0.04 are not uncommon, but the norm. 
   Right:($\tilde{v}$, $s_{2D}$) Binning strategy of Banik et al. (2024). Notice the narrow bins of $\tilde{v}$ width 0.08 in the critical $\tilde{v}<1.6$
   region. As these bins are narrower than twice the typical errors in this region, observational errors on the comparison of models to data should not be ignored,
   although they { were} ignored in Banik et al. (2024).
 }
 \end{figure*}

The analyses of C23 and C24a are further complemented by C24b's analysis using the normalized dimensionless velocity $\tilde{v}$ as a convenient dimensionless
parameter, as generally used. This parameter is intermediate between the deprojected 3D acceleration (or velocity) and the unnormalized velocity $v_{2D}$.
The relation between measured and Newtonian acceleration, if projected on the plane of the sky, reduces to the relation between $\tilde{v}$ and $s_{2D}$ normalized
by the MOND radius $r_{\rm{M}} \equiv {GM/a_0}$. C24b's work was originally motivated by the need to precisely address B24's main criticism of C23 with a data
quality cut based on the uncertainties of $\tilde{v}$. However, an analysis of the $\tilde{v}$ vs $s_{2D}/r_{\rm{M}}$ relation is interesting in its own right. 

C24b shows that the $\tilde{v}$-$s_{2D}/r_{\rm{M}}$ relation inferred from the {\it Gaia} data is fully consistent with the results of C23 and C24a. More
importantly, even when an artificial cut based on a constraint on $\tilde{v}$ uncertainties is imposed, the {\it Gaia} relation differs from the corresponding
Newtonian prediction confirming the gravitational anomaly at low acceleration. Notice also that the de-projection into 3D space applied in the studies
summarised in this sub-section, implies a careful error propagation analysis, which was included. Also, this step is absent from all the work presented in the
previous sub-section, and marks another point of independence between both sets of results, which hence become complementary in reinforcing each other through
the full accordance of their conclusions.


In summary, the six independent and highly distinct studies described above suggest that a low acceleration validity threshold for Newtonian gravity has been found,
and given the low velocity regime being probed, reveals that just as a high energy validity limit for General Relativity is implied by the inconsistency of that theory
with Quantum Mechanics, a low acceleration validity limit for it also exists, coinciding with the expectations and predictions of MOND, as first proposed in Milgrom (1983).
Moreover, the five more recent of these studies over the span 2023 - 2024 (coinciding with the use of {\it Gaia} DR3) indicate that the magnitude and trend of the low
acceleration gravitational anomaly agree with the predictions of MOND gravity theories such as AQUAL and QUMOND under the external field effect of the Milky Way. 
Notice that although the MOND radius of 2$M_{\odot}$ total mass binaries is of about 7 kau, even though typical masses of the binaries in the studies above are slightly
lower at close to 1.6$M_{\odot}$ (Hernandez group,  1.0-1.4 $M_{\odot}$ Chae Group), taking spiral galaxy rotation curves as analogies, the appearance of the first traces
of a gravitational anomaly is expected in MOND at acceleration of about 2-3 $a_{0}$, in cosistency with the anomalies reported by the above two groups appearing at 2-3 kau.

\section{Recent results reporting a preference for Newtonian models in the low acceleration regime}

\subsection{Pittordis \& Sutherland (2023)}

We begin with a close review of Pittordis \& Sutherland (2023), henceforth PS23, which established many of the  sample selection choices and statistical methodology
which after improving on the details, were { later} followed by B24. In PS23 a large sample of 73,159 wide binaries from the early version of {\it Gaia} DR3, EDR3 within
300 pc from the sun were chosen in the 5-20 kau $s_{2D}$ range, with mean total masses of a little below $1.5 M_{\odot}$ (see Fig. 20 in PS23) and $\tilde{v}$
median errors of 0.14. Given mean $\tilde{v}$ values of about 0.6, this implies mean final signal-to-noise ratios on the fitted quantities of only 4.3. Notice that
data points with much larger uncertainties were also included, at the 80 percentile level, $\tilde{v}$ errors reached 0.26, barely a 2.3 signal-to-noise ratio, Table 2
in PS23. The sample was cleaned from nearby perturbers by requiring that no other {\it Gaia} sources appeared within an angular separation of 2/3 of the wide binary
separation. Notice that this is significantly less strict than the criterion followed in H22, where the corresponding number was not 2/3, but a minimum of 8. Also, no
attempt was made at removing hidden tertiaries, which in contrast are { later} modelled. This lax selection criteria and large sample distance resulted in a much larger
sample than any of the ones described previously, at the price of using lower quality data and having less control over the systematics of the problem.

Before comparing observed $\tilde{v}$ distributions to data, the detailed effects of hidden tertiaries are calculated using a large library of orbital simulations
first developed in Pittordis \& Sutherland (2019), for both Newtonian and MOND scenarios, this last taken from the work of Banik \& Zhao (2018). Then, models are
constructed for wide binaries using either Newtonian gravity or a particular MOND variant, to which a fraction of hidden tertiaries can be added, as well as a level
of flyby contaminants, which were also knowingly not removed from the sample. The final comparison between data and models is carried out through a counts-in-cells
comparison in the $(\tilde{v}$, $s_{2D})$ plane. Only four slices are used in $s_{2D}$, and for each of these, between 50 and 70 $\tilde{v}$ cells are chosen
depending on the $s_{2D}$ range of the slice, such that the width of each cell in $\tilde{v}$ space was taken fixed at 0.1. The total number of cells for the model
to data comparison is of 250, with a resulting average number of data points per cell of 293.

Then, for both the Newtonian and the MOND scenario, optimal combinations of pure binaries, hidden tertiaries and flybys are sought so as to best reproduce the
observed distribution of $\tilde{v}$ values in the 250 cells in the ($\tilde{v}$, $s_{2D}$) plane, and a final $\chi^{2}$ goodness of fit is assigned per $s_{2D}$
slice, for three different plausible assumptions on the inherent distribution of ellipticities for each gravity model. It is important to appreciate that out of the
various effects and processes present in the observed sample, one was not included at all in the modeling: the presence of observational noise. Noise-free models
are fitted to a reality which includes substantial levels of observational noise. Since median and 80 percentile $\tilde{v}$ errors of 0.14 and 0.26
were present in the data, and since the fixed width of cells in $\tilde{v}$ space was of only 0.1, it is clear that the observed numbers of wide binaries per cell
can not be considered a pristine reflection of a physical situation, but necessarily reflect an underlying reality, plus a substantial broadening and smoothing of the
initial distribution due to observational noise.

The observed $\tilde{v}$ distributions are similar to the ones obtained for the cleaner samples of the studies mentioned in the previous section, with a peak
close to $\tilde{v}=0.6$, but this time show the addition of an extended tail much above the values expected in Newtonian or MOND scenarios for this quantity
of $\sqrt{2}$ and $1.4 \sqrt{2}$, respectively. This extended tail reaches values of up to $\tilde{v}=7$, and constitutes mostly flyby contamination. The
presence of hidden tertiaries broadens the expected $\tilde{v}$ distribution towards high values, as already mentioned, up to values of close to $\tilde{v}=3$,
see Fig. 10 in B24. Thus, the extended tail in  $\tilde{v}$ values seen in PS23 represents the presence of processes other than the one which is the subject
of the test, and a level of contamination present throughout, even at smaller $\tilde{v}$ values where the main differences between Newtonian and MOND models
appear.

The final results in PS23 are presented as a comparison of $\chi^{2}$ values for Newtonian and MOND models, over the four $s_{2D}$ slices considered, showing a marked
preference for Newtonian models over MOND ones, for all ellipticity distributions considered, in Fig 16. Three shortcomings of PS23 explain the inconsistency
of their results with those of the studies mentioned in the previous section. First, given the mean total masses are the same as in the works summarised in the
previous section, the transition from Newtonian behaviour to a MOND one will occur at the same value of $s_{2D}=2$ kau. This transition is completely missed in PS23,
as the range of projected separations covered is $5<s_{2D}/kau<20$. This does not in itself invalidate the study, it is certainly possible to explore optimal gravity
models only on one side of the transition found by the works described in the previous section. However, not including a region where all gravity models under consideration
yield the same result means that an internal consistency check on the methods used is absent.

Second, the statistical comparison performed is flawed, as noise-free
models are compared to a reality where the presence of observational noise is an important effect. As mentioned previously, median observational errors on individual
$\tilde{v}$ values are of 0.14, with errors of 0.26 and 0.34 present at the 80 and 90 percentile levels, Table 2 in PS23. Thus, it is clear that the
final membership values per 0.1 wide $\tilde{v}$ cells of the observational data used can not be considered as a result of only gravity on the wide binaries, the presence
of flybys and the presence of hidden tertiaries, the only three processes included in the models compared to the data. At an average 293 data points per bin,
a substantial variance is expected, as explicitly shown to happen in the two $s_{2D}$ bins considered in H24, with total occupancies of 108 and 466 points in the low and
high acceleration regions, respectively. Hence, a single model can not be considered as unique in terms of predicted numbers in the final 250 $(s_{2D}, \tilde{v})$
cells used for the comparison to the data. Taking for example a Newtonian model, with a thermal ellipticity distribution, a 50 \% hidden tertiary fraction and a 10 \%
flyby fraction, will not result in a unique number of points in each of these  250 cells, a variance due to the substantial observation errors should have
been included, e.g. by resampling the observed data points within their calculated errors, so as to gauge the level of variance expected in the results. Thus, the thin
$\chi^{2}$ lines shown in Fig. 16 in PS23 should more correctly be bands, and any preference of a certain model over another should take account of the range of $\chi^{2}$
values resulting from a fair comparison of models to a noisy reality.

 \begin{figure}
 \vskip 0pt
 \hskip -5pt
 \includegraphics[height=7.0cm,width=8.7cm]{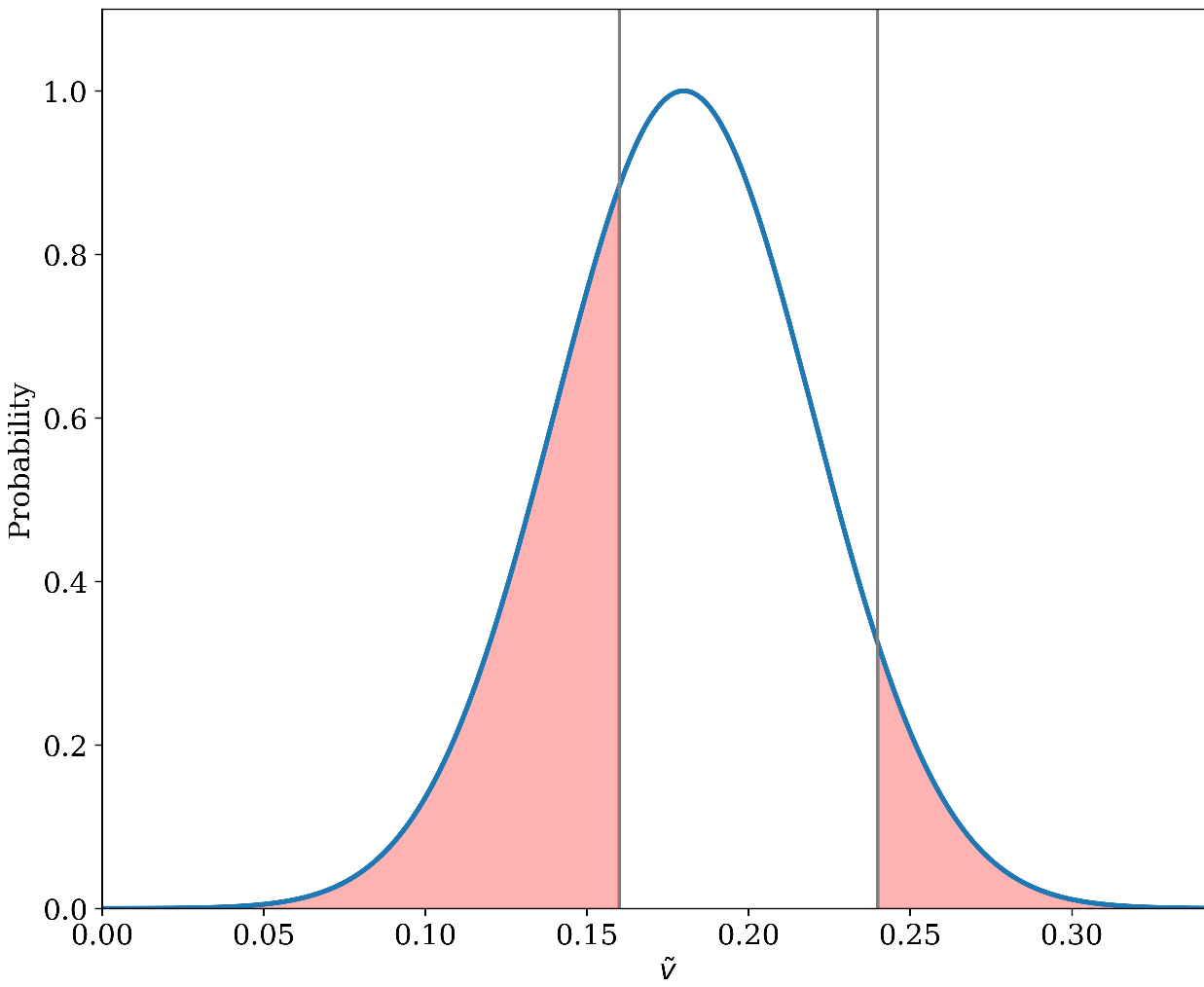}
 \caption{The figure shows the probability density function for the true position of a wide binary with inferred $\tilde{v}=0.18$ assigned to the velocity bin
   $0.16<\tilde{v}<0.24$ in Banik et al. (2024), having a less than average $\tilde{v}$ error of $\sigma_{\tilde{v}}=0.04$. In 37.5\% of such cases the
   bin identification will be erroneous. For a similar case with $\sigma_{\tilde{v}}=0.08$, still below the $\sigma_{\tilde{v}}=0.1$ upper quality
   cut relevant for this region, the probability the assigned bin is off by one is of 48.2\%, while the probability that the bin has been miss-assigned
   by two steps is of 14.6\%. This is serious, as observational errors are not considered in the comparison between models and observations by those authors.}
 \end{figure}

This last point implies not only a reassessment of the degree to which the data presented prefer a Newtonian model over a MOND one, but also, a change in the preferred
model, for the reasons explained below. As explained previously, the addition of noise (not included in the models of PS23 but present in the data used) implies a
fairly symmetrical broadening of the assumed input model $\tilde{v}$ distribution, as is general to the introduction of any Gaussian observational noise component.
The introduction of
hidden tertiaries, as explained in the previous section, is asymmetrical, in producing a broadening almost exclusively towards larger values, not smaller ones.
Hence a MOND reality to which noise is added, will present a symmetrical broadened distribution of $\tilde{v}$ values, in comparison to the noise-free underlying
input. When comparing to a noise-free MOND model to which hidden tertiaries are added, the extension towards small $\tilde{v}$ values produced by the presence of
noise will appear as an excess which the noise-free MOND model will be unable to match. A Newtonian noise-free model however, will have no such problem, as it
is intrinsically shifted to smaller values in comparison to the MOND one. Indeed, this is clearly apparent in the comparisons of {\it Gaia} data to MOND models
in figures 13-15 in PS23, the data always have an unmatched excess at small $\tilde{v}$ values. This represent the extension towards the left due to the present but
un-acknowledged effect of noise. In the corresponding Newtonian figures, 10-12 in PS23, no such unmatched excess appears at small values. Notice that this good match is
actually the indication of an inconsistency of the Newtonian models to the data presented, as these Newtonian models do not incorporate the presence of noise, present at
a median level of 0.14 in $\tilde{v}$, with the noisiest 20 percentile actually having errors above 0.26. If such noise were added to the Newtonian curves shown, it is
clear that they would all over-shoot the data at small values of $\tilde{v}<0.5$. Towards the falling edge of the main distribution, Newtonian noise-free models match a
noisy MOND reality, through the addition of a suitable amount of hidden tertiaries and flybys, which broaden the final model distribution towards larger values as much
as required to match the data. In summary, the comparison of noise-free models to a reality which incorporates a non-negligible noise component biases the counts-in-cells
approach towards a model with a lower effective gravity. 

A final problem with PS23 is the lack of a full internal check of the method using simulated data sets. Recall that in C23, C24a, C24b and H24, full
synthetic data sampling and reconstruction experiments were conducted to asses the internal consistency of the statistical methods used, which were
thus tested and shown to be free of biases, at least regarding the physical ingredients modelled, which always included full accounting of observational noise.
As mentioned above, the direct comparison of noise-free models to data which include observational noise, biases the results of PS23 towards a lower effective
gravity model in the $s_{2D}$ range probed, as the broadening which in the observed samples is produced by symmetric observational noise, in the theoretical models
is produced by the addition of hidden tertiaries, an asymmetric effect. Had a deep Newtonian region been included, the problem would have been identified, as the best
fit gravity model in the $s_{2D}<2000$ kau region would have resulted sub-Newtonian (having an effective gravitational constant below the standard G).

\subsection{Banik et al. (2024)}

In B24 the authors of PS23 colaborate in a further wide binary gravity test, using this time the latest {\it Gaia} DR3. Using a medium-sized sample of 8,611 wide
binaries within 250 pc from the Sun with $2000<(s_{2D}/au)<30,000$, to attempt a test of gravity in this separation range. Binaries were required to have velocities
along the line of sight only for one component, rather than both, as in H22, H23 and H24. Careful mass inferences are included,
together with detailed error propagation on the reported {\it Gaia} parameters to yield both rigorous $\tilde{v}$ and $s_{2D}$ values for the wide binaries studied,
and accurate 1$\sigma$ errors on the dimensionless $\tilde{v}$ data to be used, $\sigma_{\tilde{v}}$. These $\tilde{v}$ errors are then used to define the sample, which
is selected to have good quality data, with final $1\sigma$ errors on the $\tilde{v}$ values which are restricted to being below 0.1 max (1, $\tilde{v}$/2).
i.e. for small values of $\tilde{v}$ a fixed upper bound of 0.1 in the error is imposed, while for large values of $\tilde{v}$ the upper limit is introduced on the resulting
signal-to-noise ratios on this quantity. The above choices are motivated by the critical nature of the $0<\tilde{v}<2$ region in discriminating between
Newtonian and MOND scenarios, and ensure that the error on the $\tilde{v}$ values used is at most equal to the difference of 20\% expected between a Newtonian
and a MOND model near the peak of the  $\tilde{v}$ distribution, which occurs at close to $\tilde{v}=0.5$. As a result, measurement errors in B24 can be expected to
broaden the distribution from 0.5 to $\sqrt(0.5^2 + 0.1^2) = 0.51$, which represents only one tenth of the 20\% shift that is expected in MOND. The broadening of the peak
will hence be minor, although as we shall see, the variance due to this errors will prove to be a dominant effect, which was ignored in B24.

Then, a 2 dimensional 540 cell binning in the ($s_{2D}$, $\tilde{v}$) plane is imposed, with a binning strategy which was previously established and decided upon to
optimise the possibility of distinguishing between Newtonian and MOND scenarios, Banik et al. (2021). This binning of the ($\tilde{v}$, $s_{2D}$) plane is then
populated using the carefully inferred observational $\tilde{v}$ values described above. However, the corresponding $\sigma_{\tilde{v}}$ values play no 
role in the gravity inference procedure. Their role is restricted to defining the sample, as can be confirmed from the first paragraph in Section 2.3 and
from section 2.4.6 in B24. Once this ($\tilde{v}$, $s_{2D}$) plane has been populated, it serves as a unique template against which various models
are compared and evaluated through a counts-in-cells procedure. In going from the 73,159 binaries and 250 cells of PS23 to 8,611 binaries and 540 cells
in B24, the average occupancy numbers fell from 292.6 in PS23 to only 16 observed binaries per cell in B24, as shown later, this introduces a very large
variance on the final inferred results, which was ignored in B24.

Each model is defined through various parameters including the $\alpha$ parameter mentioned
in the previous section in connection with the probability distribution of ellipticities, the $f_{\rm{multi}}$ parameter also mentioned previously in connection
with an assumed fraction of hidden tertiaries, and an $\alpha_{grav}$ parameter which is defined as a sliding index between Newtonian gravity and MOND,
with 0 corresponding to Newtonian gravity and 1 to a particular MOND model, QUMOND (Milgrom 2010). Further parameters are also introduced to describe
the statistical distribution of hidden tertiary orbital parameters. A simultaneous fit to seven parameters describing the hidden tertiary frequency and
orbital parameter distributions, an ellipticity probability function, a fraction of line-of-sight contaminants and potential changes in gravity is then
implemented by these authors, as they are dealing with a sample where no strict kinematic contaminant exclusion strategies have been followed, and where
no deep Newtonian region is employed to calibrate the hidden tertiary fraction.

Finally, a MCMC method is used to recover optimal parameter values, through counts-in-cells binomial comparisons of models to the observational
binned ($s_{2D}$,  $\tilde{v}$) plane described above, which is treated as a fixed, unique template. A final goodness of fit probability is assigned to each model
as defined through a log-likelihood value for a model given the occupancy numbers of each 540 pixels:

\begin{equation}
  ln P = \sum_{Pixels} ln P_{pixel}
\end{equation}

where

\begin{equation}
P_{pixel} = \frac {N!}{(N-k)!k!} p^{k} (1-p)^{N-k}.
\end{equation}

In the above equations $p$ is the fraction of the total number of observed binaries lying within a certain cell, $k$ the resulting binaries
from the model being tested, and N the total sample size.

As happens in PS23, the consistency of bin size and data errors was neglected. The approach presented would be rigorous if the number of
observed binaries per ($s_{2D}$, $\tilde{v}$) bin could be treated as being accurately determined. This last condition would apply if and only if,
the errors on both $\tilde{v}$ and $s_{2D}$ were always much smaller than twice the bin sizes. Whilst this condition is easily met by the
$s_{2D}$ values and their errors from the exquisite {\it Gaia} catalogue, it is failed by the ($\tilde{v}$, $\sigma_{\tilde{v}}$)
values being used. This can be readily confirmed by looking at the distribution of  $\sigma_{\tilde{v}}$ values from B24, shown in their
Fig. 4, which we reproduce in the left panel of our Fig. 1, and from their binning of the ($\tilde{v}$, $s_{2D}$) plane, their Fig. 6, which we reproduce
in the right panel of our Fig. 1. From the binning of the ($\tilde{v}$, $s_{2D}$) plane we see that the sensitive $0<\tilde{v}<1.6$ region of the plane,
where most discriminating power between a Newtonian and a MOND model resides, has been split into 20 bins, resulting in $\tilde{v}$ bins of width 0.08.

 \begin{figure*}
 \vskip 0pt
 \hskip -5pt
 \includegraphics[height=7.0cm,width=8.5cm]{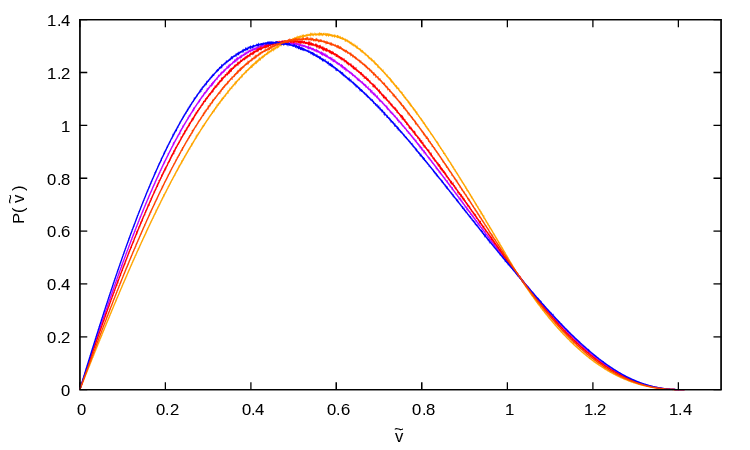}
 \includegraphics[height=7.0cm,width=8.5cm]{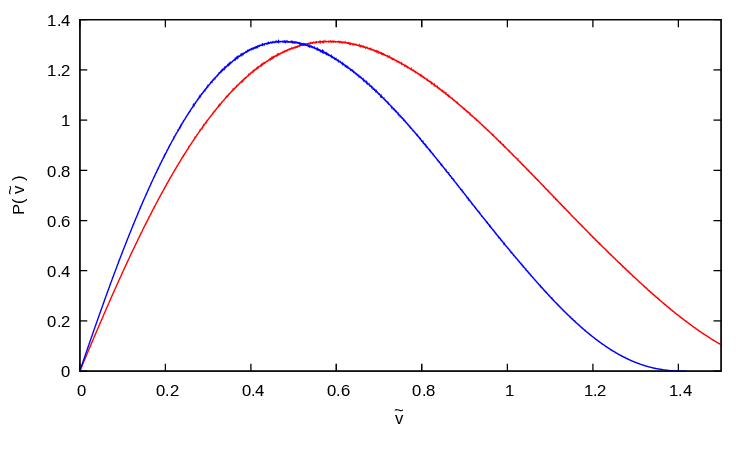}
 \caption{Left: Various $P(\tilde{v})$ curves for different assumed ellipticity distributions parameterised using $P(e)=(1+\alpha)e^{\alpha}$: $\alpha=0.6$,
   yellow, $\alpha=0.8$, orange, $\alpha=1.0$, thermal, red, $\alpha=1.2$, purple and $\alpha=1.4$, blue, respectively, assuming Newtonian gravity.
   Right: Comparison of $\tilde{v}$ distributions with the same fixed $\alpha=1.2$ but for $G \to \gamma G$, a Newtonian model at $\gamma=1.0$ and a MOND
   one at $\gamma=1.512$, blue and red curves respectively. Notice the total offset in the peak of the curves in the left panel approximately corresponds to that
   of the curves in the right one, showing that any inference of $\gamma$ in the $1.0<\gamma<1.5$ range using as input full $\tilde{v}$ distributions will
   necessarily be correlated with assumed or inferred values of $\alpha$. }
 \end{figure*}

We can now check from their Fig. 4 that in the corresponding $0<\tilde{v}<1$ and $1<\tilde{v}<2$ ranges, about 2/3 and 4/5 of cases respectively, have
errors larger than half the bin size which in this $\tilde{v}$ ranges is of only 0.04. If we take an observed binary sitting in the centre of one
of these bins and having a more than typical $\tilde{v}$ error of $\sigma_{\tilde{v}}=0.04$, the probability that it should actually reside in either the
bin above or the bin below is calculated as the fraction of the integral of a Gaussian lying { outside} a 1$\sigma$ interval, 31.7\%. If we take a more
typical case of a binary lying half-way between the centre of one of this $\tilde{v}$ bins of width 0.08 and the edge, and having the same more than
typical $\sigma_{\tilde{v}}=0.04$ value, the probability of it having been scattered by observational errors from one of the neighbouring $\tilde{v}$ bins
raises to almost 40\% at 37.5\%, as shown in our Fig. 2. The situation clearly worsens for larger errors. Notice from the left panel of Fig. 1 that more
than half of the binaries used by Banik et al. (2024) have  $\tilde{v}$ errors between 0.04 and 0.1, for  $\tilde{v}$ values below 2. In fact, values of
$\sigma_{\tilde{v}}=0.08$, are not rare, and lie below the upper quality cut of $\sigma_{\tilde{v}}<0.1$ for the $\tilde{v}<1.6$ region where the bin
size is of 0.08. When considering such $\sigma_{\tilde{v}}=0.08$ cases, the probability that the observed binary has been erroneously assigned to a
certain bin from one of the contiguous ones raises to 61.7\% and 62.8\% for the two locations within the bin discussed above. Indeed, for this last
$\sigma_{\tilde{v}}=0.08$ value, still within the upper limit of 0.1 imposed, the probability that a wide binary lying half-way between the centre of one
of { these} bins and the edge has been miss-assigned by two bins rather than just one, is already of 14.6\%. The mean value of $\tilde{v}$ shown in the figure
is of only 0.18, smaller than the mean values which are of close to 0.6. However, the argument made above depends exclusively on a comparison of error size
to bin size and nothing more, and hence applies equally to any other central $\tilde{v}$ value closer to the median value.

It is hence obvious that the observational template used by the authors in B24 can not be assumed as being error free, at the fine
$\tilde{v}$ binning imposed, there has necessarily been a non-negligible blurring due to the observational noise present in the data. This should have
been dealt with by either taking wider bins, tighter $\sigma_{\tilde{v}}$ limits, or including MC re-samplings of the data within their inferred confidence
intervals. None of these standard strategies for accounting for the presence of noise in discrete binned observations-to-model comparisons (e.g. Hernandez et al. 2000)
were followed, and in B24 the unavoidable blurring of their fine $\tilde{v}$ binning strategy by the comparatively large $\sigma_{\tilde{v}}$ they carefully obtained,
is ignored. It is an easily verifiable fact that the optimal parameter estimation, and corresponding statistical significance level estimation, method implemented
by B24 does not include any consideration of the errors in the data whatsoever, $\sigma_{\tilde{v}}$ values are used exclusively in defining the sample, and ignored once
this first step has been completed. Contrary to standard practice, observational errors are not even added to the models constructed from the various parameter choices
before comparing to the observational template. Models are compared directly to the observational template without including a blurring of the model $\tilde{v}$ values
compatible with the observations being used. Therefore, the problem includes a non-acknowledged observational noise component.

In summary, B24 presents an update on PS23, with the use of the latest
{\it Gaia} DR3 catalogue rather than the previous EDR3, a substantially detailed estimation of full observational errors, a more detailed and careful
hidden tertiary modelling both under Newtonian and MOND models, a sophisticated model optimisation algorithm to optimally identify best fit parameters and
a binomial likelihood function for estimating the goodness of fit of the models considered in comparison to the data. However, the problems of PS23 persist,
although the $s_{2D}$ range was extended downwards from 5kau to 2 kau, it still excludes the deep Newtonian region where hidden tertiary fraction calibrations
and internal consistency checks on the gravity inference procedure could have been performed. The similarity in the sample selection of B24 and PS23 extends to the
isolation criteria, which again in B23, only excludes from consideration wide binaries having dim companions closer to 2/3 the binary separation. Thus nearby
perturbers at 5, 3, or even 1 times the binary separation are accepted, and their effects not included in the modelling.

With 540 bins in the $(s_{2D},\tilde{v})$ plane, final average occupancy numbers of only 16 result. Hence, it is clear that a significant variance is implied by the
random shifting due to the level of observational noise present. As in PS23, this variance is ignored and comparisons to a single observational template are presented
as single values in the goodness of fit parameter found for the models compared, which more rigorously, should include a confidence interval on the goodness of fit
parameters found for each model.

Also, still no simulations using synthetic data were included to check the overall performance of the methodology presented, and no inclusion of observational errors in
the statistical procedure used to infer the law of gravity were included. Although the ratio between median $\tilde{v}$ errors and bin size in the critical region
was reduced from PS23 (0.14/0.1=1.4) to 0.06/0.08=0.75, this ratio still remains sufficiently high to imply a significant variance in the assigning of a goodness-of-fit
parameter to a model-to-data comparison, which was not estimated. Indeed, the actual upper error limit in B24 for the sensitive $\tilde{v}<1.6$ region in B24 is of 0.1,
even larger than the $\tilde{v}$ bin size in that region, of 0.08. A full simulation using the same binning scheme of B24 is included in Section 5, where observational
error sampling examples illustrating explicitly the level of variance intrinsic to the method used in B24 are developed.

Further, higher than directly inferred best-fit $f_{\rm{multi}}$ fractions result. Indeed, the inferred $f_{\rm{multi}}$ obtained by B24 is of 69\% in their best-fit Newtonian
model, although certain parameter variations tested in B24 can bring down this value, at the cost of a poorer Newtonian fit. This value for the optimal Newtonian model
is in conflict with direct observational estimates of this quantity in local wide binaries, which are well established to lie below 50\%, e.g.
Moe \& Di Stefano (2017). Notice that the careful $f_{\rm{multi}}$ calibration of C23 using the Newtonian high acceleration region returns in his case values in the range
$0.25 \lesssim f_{\rm{multi}} \lesssim 0.5$, fully consistent with independent direct estimates for this parameter.

The authors in B24 are aware of the fact that their best fit recovered $f_{\rm{multi}}$ fraction is incompatible with independent determinations, as also happens with
their recovered ellipticity distribution index of $\alpha >1.85$, which has been observationally determined (Hwang et al 2022) to lie in the range
$1<\alpha<1.5$ for the {\it Gaia} wide binaries in question, see their Table 2. However, this is not presented as a problem in B24 as their inferred models of
gravity, their $a_{grav}$ parameters, appear totally uncorrelated with either of these two extra inferences. As we shall see in the following section,
it is precisely this absence of correlation between parameters of the problem which are linked at a fundamental physical level, which
once again alerts to the dominance of an unrecognised observational noise component in B24. We have attempted a clear review of B24, where the presentation is sometimes
not easy to follow, due to the extensive nature of that paper and the lack of clear signposting presentation strategy.

\section{Posterior scalings in the results of B24}

At a fundamental level, the wide binary gravity test reduces to comparing the instantaneous relative velocities and separations between members
of binary systems against predictions from various gravity models. It is unavoidable that any particular
relative velocity value observed at a given separation, can be interpreted through an infinity of combinations of orbital ellipticity, total binary mass
and effective gravitational strength. The presence of a hidden tertiary alters the total mass of the system in comparison to its inferred value when
assuming both observed stars are individual stars, plus introduces a direct kinematic contaminant due to the additional orbital velocity of the close
binary, which generally boost the relative velocities of wide binaries, particularly at the widest separations where intrinsic velocities are lower.
Thus, a given observed wide binary relative velocity value can result from either an enhanced effective gravitational constant, or the presence of a
hidden tertiary. Hence inferences of the effective value of G will necessarily anti-correlate with the assumed hidden tertiary fraction
e.g. Clarke (2020).

A similar situation arises when considering changes to the assumed ellipticity distribution for a population of wide binaries, relative velocities will change
for different assumed ellipticity distributions, and hence, inferred values of effective gravitational strength will correlate with the particular
ellipticity distributions assumed. The left panel of Fig. 3 shows $\tilde{v}$ distributions for a Newtonian model, for different values of the $\alpha$
parameter specifying the ellipticity distribution mentioned in the previous section. We see a clear shift in the position of the peak of the $\tilde{v}$
distribution as a result of changing the details of the ellipticity distribution assumed. For comparison, the right panel of Fig. 3 shows two $\tilde{v}$
distributions at a fixed ellipticity distribution of $\alpha=1.2$, for both a Newtonian $\gamma=1.0$ and a MOND $\gamma=1.512$ model. The shift in the
peak of the resulting distributions is of comparable magnitude to what results from changing $\alpha$ within observationally constrained ranges,
at fixed $\gamma$. This implies that any inference on the effective strength of gravity based on $\tilde{v}$ distributions will be correlated to the
assumed or inferred details of the ellipticity distribution of the wide binary sample being treated. Indeed, in e.g. Pittordis \& Sutherland (2018),
PS23 or H24, clear correlated scalings in the inference of effective gravity and details of assumed ellipticity distributions are readily apparent.

We now turn to the nominal results of B24. The bottom row of their Fig. 14 gives the correlations between the inferred gravity parameter
and the recovered values of the other parameters, which we reproduce here in Fig. 4. The first two parameters refer to the orbital details of the wide
binaries, the third is actually the ellipticity distribution index $\alpha$ mentioned here, the fourth the hidden tertiary fraction, the fifth a parameter
relevant to describing the orbital details of the hidden tertiary orbits, and the sixth, the assumed fraction of line-of-sight contamination of a model.
It is remarkable that neither the third, nor the fourth nor the fifth nor the sixth parameters show any significant correlation with the inferred
$\alpha_{grav}$ parameter of the models, the effective gravity inference presented appears independent of the recovered details of the orbital eccentricity
distribution, the inferred hidden tertiary fraction, the level of line-of-sight contamination or the details of the hidden tertiary orbits. Best fit results
are given by the red curves, while blue curves show results for a variant model where the ellipticity is forcibly constrained to its directly determined
ranges from Hwang et al. (2022), even imposing this change in the ellipticity distribution results in no changes in the recovered gravity model.

The presence of fundamental correlations between the parameters of a problem is generally taken
as a quality test of the sample selection and inference procedure implemented. In this case, such fundamental physical scalings are essentially absent.
This is quite probably the result of the fact that B24 are working with a sample where the presence of an important observational noise
component has been ignored and thus, the inference procedure has been diluted to the point that clear physical scalings of the problem have been erased.
Breaking some of the degeneracies between the many parameters being simultaneously fitted would necessarily have required the inclusion of independent
data, such as the use of the angle between $v_{2D}$ and $s_{2D}$ which Hwang et al. (2022) use to estimate the case-by-case ellipticities which
C23, C24a and C24b use in their inferences.

 \begin{figure}
 \vskip 0pt
 \hskip -5pt
 \includegraphics[height=2.5cm,width=8.5cm]{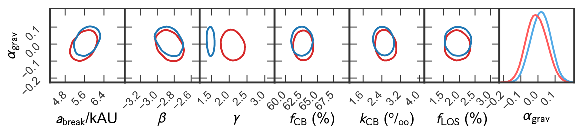}
 \caption{Correlations between inferred gravity parameter, $\alpha_{grav}$, and the other parameters of the problem, bottom row of Fig. 14
   in Banik et al. (2024). Notice the almost complete lack of correlation between the recovered $\alpha_{grav}$ and the fraction and orbital
   characteristics of the assumed dominant hidden tertiaries, fourth and fifth parameters, and the equal lack of correlation between $\alpha_{grav}$
   and the assumed ellipticity distribution, $\gamma$ in the above plot, corresponding to $\alpha$ in the text. Best fit B24 results are given by
   the red curves, while blue curves show results for a variant model where the ellipticity is forcibly constrained to its directly determined
   ranges from Hwang et al. (2022), see text.
 }
 \end{figure}

Further internal inconsistencies appear in the results of B24 in their figure 12, which summarises the comparison between their best
fit Newtonian and MOND models for four $s_{2D}$ ranges. It is a fundamental property of MOND that by construction, at high accelerations the description
of gravity recovers Newtonian gravity exactly, and that the trend towards Newtonian gravity with acceleration is monotonic. High accelerations
will necessarily occur towards low $s_{2D}$ values. It is therefore hard to understand the fact that of the four $s_{2D}$ ranges shown in Fig. 12 in
B24, the first panel, the one showing the tightest binaries at $2<(s_{2D}/kau)<3$, is the one where the MOND and Newtonian models
differ the most between them. By contrast, all results of the Hernandez and Chae groups unequivocally show a monotonic convergence towards the Newtonian
model on approaching high accelerations, with the $2<(s_{2D}/kau)<3$ region showing always a close convergence of Newtonian and MOND predictions.

Figure 12 in B24 also illustrates the difficulty in dealing with a sample where no kinematic contaminants have been excluded or independently
constrained. From Fig. 10 in that study it is obvious that pure wide binaries end before $\tilde{v}=1.8$ values, so that the extensive distributions
observed beyond this limit in their Fig. 12 are due to kinematic contaminants. This alerts to the fact that that same level of kinematic contaminants will
also be present in the crucial $0<\tilde{v}<1.5$ region, clearly a contribution comparable or larger than the difference between the models being tested,
as seen in their Fig. 12. This last point applies also to the error bars shown, either the level of kinematic contaminants present, or the internal error bars of
the study due to the presence of observational noise are each comparable to the difference between the models being compared. We note also that in this same
figure in B24, results for the $5<(s_{2D}/kau)<12$ range appear to favour the MOND model over the Newtonian one, although as argued in B24, this is a purely
subjective impression, as such a preference is actually limited to a small number of $\tilde{v}$ cells.  Also, the enhanced presence of
kinematic contaminants and the concern over the presence of nearby perturbers (this last not explicitly excluded beyond 2/3 the binary separation for faint
{\it Gaia} sources, or modelled in this study), bring into question their results in their fairly mixed, widest $12<(s_{2D}/kau)<30$ region, although as
argued in B24, this last $s_{2D}$ bin is of little consequence to the overall assessment. Still, { these} last points bring into question the claimed 19$\sigma$
preference of the Newtonian model over the MOND one, as even a 6$\sigma$ preference between alternatives typically appears as an extremely clear predominance
of the preferred model, significantly above the presence of contaminants or error levels. As we show in Section (5.2), accounting for the presence of internal
variance due to noise in the procedure of B24, leads to significantly different conclusions to those presented in B24.

We remark that the value of a scientific result is not defined exclusively by its formal statistical confidence level, but also by the degree to which
systematics have been controlled and accounted for. For this in H22, H23, C24a and H24, extreme care was taken to
construct samples highly free of kinematic contaminants which then do not have to be modelled simultaneously, including accounting for their intrinsic physical
correlations on the final results. In C23, the only kinematic contaminant not explicitly excluded, hidden tertiaries, is calibrated accurately
using a deep Newtonian sample in the $0.2<(s_{2D}/kau)<2$ regime. The presence of a deep Newtonian region in all our papers serves also as a consistency check
on the entire procedures being implemented, as it becomes a region where all scalings and values can be checked. This consistency check is absent from
B24 which is restricted to the $2<(s_{2D}/kau)<30$ range. Indeed, had a deep Newtonian $0.2<(s_{2D}/kau)<2$ regime been included in B24, a discontinuity
in the best fit parameters across the $s_{2D}=2$ kau divide would have signalled the presence of a regime change.  

\begin{figure*}
  \centering
  \includegraphics[width=0.9\linewidth]{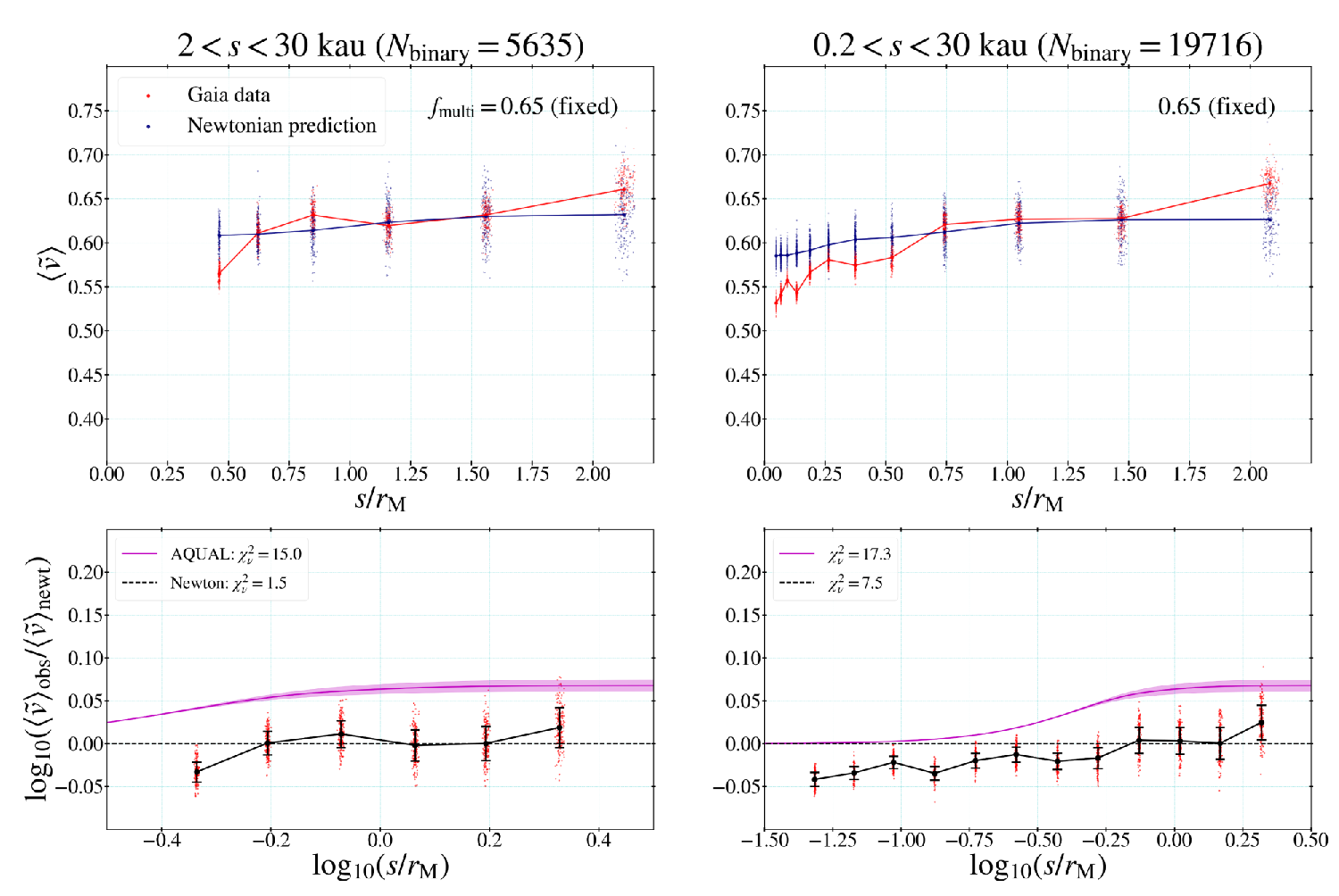}
    \vspace{-0.3truecm}
    \caption{Left: This figure compares the observed (C24b) binned medians of $\tilde{v}$ with the corresponding Newtonian predictions for a sample within the narrow
      range $2<s_{2D}<30$ kau used by B24. A deliberately high value of $f_{\rm{multi}}=0.65$ for the Newtonian model is used to be consistent with B24. In the upper panel,
      the red and blue points represent distributions of medians from 200 MC results (see C24b for the details). The bottom panel compares the ratio
      $\langle\tilde{v}\rangle_{\rm{obs}}/\langle\tilde{v}\rangle_{\rm{newt}}$ with the Newtonian and AQUAL predictions. As the values of the reduced chi-squared
      ($\chi^2_\nu$) indicate, the Newtonian model is consistent with the data while the AQUAL model appears ruled out. This conclusion with $f_{\rm{multi}}=0.65$
      is consistent with that of B24. Right: A wide range of $0.2<s_{2D}<30$ kau used by C23 is considered. With $f_{\rm{multi}}=0.65$, at high accelerations, the
      Newtonian data is now inconsistent with the Newtonian prediction as a consequence of the artificially high value of $f_{\rm{multi}}$. Once $f_{\rm{multi}}$ is
      adjusted to a lower value of about 0.4 to match the Newtonian regime data, the AQUAL model becomes consistent with the behaviour of
      $\langle\tilde{v}\rangle_{\rm{obs}}/\langle\tilde{v}\rangle_{\rm{newt}}$ while the Newtonian model is ruled out (see C24b for the details). When working with
      an excessively large  $f_{\rm{multi}}$ value, an artificially low effective $\gamma$ results, however even in this case, the regime transition is evident,
    if covered by the range of data considered, which is not the case in PS23 or B24.}
 \end{figure*}

Also, the robustness of our conclusions was thoroughly tested in all of both the Chae and the Hernandez publications, exploring changes in the sample
selection strategies to verify our inferences had converged with respect to the systematics not explicitly included in the various statistical methods
employed. In contrast, B24 consider only data samples from which the associated observational errors were ignored when
comparing to the gravitational models being probed. An exploration of a high quality sub-sample, e.g. where only binary candidates where both stars have
{\it Gaia} radial velocities, for example, could yield some comparison to a case where at least one of the systematics, unbound binaries and projection
effects, can be more reliably excluded, if using a comparison fully accounting for the inevitable presence of observational noise.

We finally note that all systematics will necessarily grow with distance, as the quality of observations and their internal systematics (not only their
formal statistical errors) also increase. In this sense, the limit distances of the samples used by the three groups discussed here are of 200 pc and
80 pc in C23, 125 pc in H23 and H234, 135 pc in H22, 200 pc in C24a and C24b, 250 pc in B24 and 300 pc in PS23.

\section{Explicit exploration of the effects of the implementation problems in PS23 and B24}

\subsection{Consequences of excluding the deep Newtonian regime and of obtaining a higher than present $f_{\rm{multi}}$}

Unlike the various studies described in Section~2, the two recent studies, PS23 and B24, used only a narrow dynamic range of low acceleration with $s_{2D}>5$ kau
in PS23 and $s_{2D}>2$ kau in B24, excluding the Newtonian regime binaries. Here we carry out a numerical experiment to explicitly show the consequence of such a
narrow dynamic range. Since B24 considered the relatively lower limit of 2 kau, we take this value here so as to clarify their results.

 \begin{figure*}
 \vskip 0pt
 \hskip -5pt
 \includegraphics[height=5.9cm,width=9.5cm]{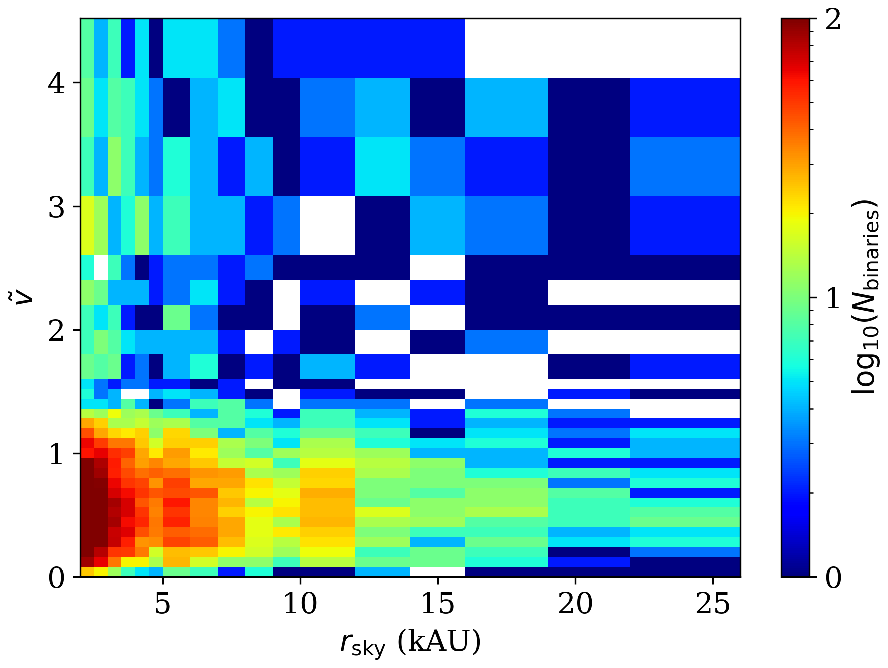}
 \hskip -9pt \includegraphics[height=5.9cm,width=8.4cm]{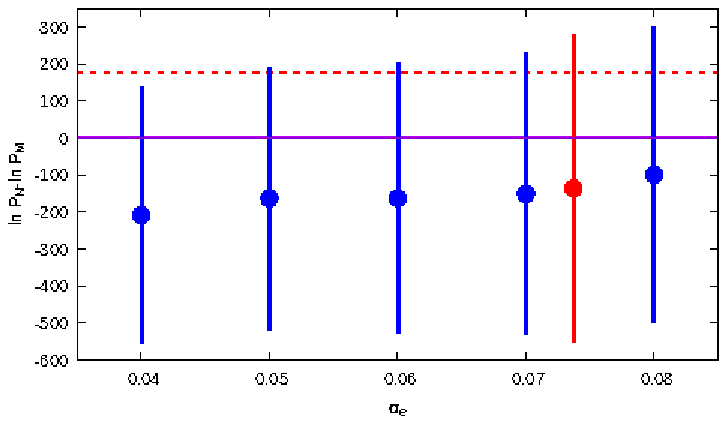}
 \caption{Left: The figure shows a counts-in-cells pixelation of the ($s_{2D},\tilde{v}$) parameter space for a 9,155 mock wide binary sample created using the MOND
   AQUAL model, including a hidden tertiary fraction as independently constrained of 0.4 and an statistical distribution of ellipticities as given by Hwang
   et al. (2022), with $1.0<\alpha<1.4$. Random Gaussian errors on $\tilde{v}$ values with amplitudes as found in the B24 study were added to the particular
   realisation shown in the figure. The pixelation scheme used is exactly the same as was used in B24.
   Right: Log likelihood values for the comparison of the best fit Newtonian and MOND noise-free models of B24 to 50 random realisations of observational noise
   addition to the mock AQUAL sample constructed, $ln P_{N}-ln P_{M}$. The blue dots with 1$\sigma$ error bars give results for the use of fixed amplitude errors of
   dispersion as given in the x-axis, while the red dot with 1$\sigma$ error bar shows results for observational error distributions taken from the {\it Gaia}
   catalogue and matching the ones used in B24. The solid horizontal line divides particular models showing a preference for the MOND model over the Newtonian, below the
   line, to those where the opposite holds, above the line. The dashed horizontal line gives the value of $(ln P_{N}-ln P_{M})=175$ for preference of the Newtonian
   model used here to the MOND model used here, reported as a unique point in B24. The substantial variance imposed by the observational errors present
   and the small average cell occupancy numbers of only 16, was ignored in B24.
 }
 \end{figure*}

B24 argued for an extremely significant preference of Newtonian gravity with $f_{\rm{multi}}\approx 0.6-0.7$ which is clearly higher than both nearby survey results
and the values obtained by C23 and C24a,b which in all cases yield $f_{\rm{multi}} < 0.5$. This is true also for the binary samples under consideration because resolved
hierarchical cases are not included in them. To see the effect of assuming an excessively high value of $f_{\rm{multi}}$, we consider a sample of 19716 binaries from C23
that satisfy fractional errors $<0.005$ on $v_{2D}$. This sample is for the $0.2<s_{2D}<30$~kau { range} within 200~pc from the Sun. The subsample with the B24 range
of $2<s_{2D}<30$~kau has 5635 binaries. 

Since B24's analyses were done exclusively using the $\tilde{v}$ variable, we consider C24b's test with the $\tilde{v}$ - $s_{2D}/r_{\rm{M}}$ relation. The reader is referred
to C24b for the details of the analysis. Although B24's sample includes kinematic contaminants such as fly-bys and binaries with close companions, the underlying
binaries must be equivalent to C23's sample if those kinematic contaminants were taken into account properly by B24. Now we consider a fixed value of
$f_{\rm{multi}}=0.65$ which is consistent with what B24 found. With this value fixed, the left column of Figure~5 compares the {\it Gaia} data with the Newtonian
prediction for the subsample with $2<s_{2D}<30$~kau. As the upper panel shows, the Newtonian model has a near perfect agreement with the {\it Gaia} data. The bottom
panel explicitly shows a reduced $\chi^2$ test with the observed-to-Newtonian ratios of binned medians of $\tilde{v}$. Clearly, the Newtonian model is highly preferred
over the AQUAL option that predicts a significant boost as represented by the magenta band. Formally, Newton is preferred over AQUAL at well above $5\sigma$. In this sense,
this result is consistent with B24's results.  

However, if we consider the full sample with $0.2<s_{2D}<30$~kau, the above conclusion is faced with a serious difficulty as shown in the right column of
Figure~5. With $f_{\rm{multi}}=0.65$, for binaries in the Newtonian regime the observed binned medians of $\tilde{v}$ are clearly lower than the Newtonian
predictions. As the bottom panel shows, the Newtonian model is assigned a $\chi_\nu=7.5$. At first glace, it appears that the Newtonian model is still relatively
preferred over the AQUAL model. However, such a relative preference is an illusion because the data at the Newtonian regime are mismatched. If we adjust
$f_{\rm{multi}}$ back to a normal value of about $0.4$ to match the Newtonian regime data, the AQUAL model now agrees well with the observed behaviour at all
$s_{2D}/r_{\rm{M}}$ ranges, while the Newtonian model does not. This experiment illustrates the importance of the Newtonian regime data excluded by B24 and PS23. See C24b
for further details. In summary, when working with an excessively large $f_{\rm{multi}}$ value, an artificially low effective $\gamma$ results, however even in this case,
the regime transition is evident, if covered by the range of data considered, which is not the case in PS23 or B24.

\subsection{Consequences of ignoring observational errors when assigning statistical goodness-of-fit values through the counts-in-cells approach of Banik et al. (2020)}

In order to gauge the confidence intervals on inferred parameters given the presence of observational noise, it is customary to perform MC resampling of the
data used within their reported errors, to construct samples of synthetic data. Each of these is then treated exactly in the same way as the original data,
and a set of inferred parameters is thus obtained. { Since noise is a stochastic process}, the inferred parameters for each of the synthetic samples will vary
somewhat. It is from the distribution of values of these inferred parameters that the confidence intervals on the parameters inferred from the original observed
data are obtained. A full description of the above can be found in, for example, the classical reference of Press et al. (2007), chapter 15.6. Several variants
of this approach exist towards inferring confidence intervals on recovered parameters from data in the presence of observational noise, but none proceeded
through ignoring the presence of observational noise in the data being treated, as was the case in PS23 and B24.

To determine the effect of this omission, we next perform an experiment to gauge directly the degree to which observational errors imply a variance on the assigned
preference of the best fit Newtonian model over the best fit MOND model in B24. We begin by taking a sample of 9155 ($s_{2D}$, $M_{T}$) values in the range
$2<s_{2D}/$kau$<20$ from the sample of C24b. To each of this $s_{2D}$ values a $\tilde{v}$ value is assigned by assuming a standard AQUAL model, following the
procedure described in C24b, (see section 3.4 in that paper). An individual gravity boost factor for each simulated wide binary was taken based on each $s_{2D}/r_{M}$
value, a low-acceleration limit boost factor of 1.37 resulted. Then, to a fraction of $f_{\rm{multi}}=0.4$ of the cases, a hidden tertiary is added
as a mass and kinematic contaminant, again following the procedures described in the previous references. The assumed ellipticities were drawn from the statistical
distribution proposed by Hwang et al. (2022) mentioned previously, using an $\alpha$ parameter taken as a function of $s_{2D}$ to match the results of Hwang et al.
(2022) for the $2<s_{2D}/$kau$<20$ range used, eq. (18) in C24a. Statistical Gaussian errors on the $\tilde{v}$ distribution resulting from the addition of the
hidden tertiary fraction described were then added either by assuming a constant $\sigma_{e}$ for them, or by taking the $\sigma_{ei}$ values from the observed wide
binary sample from C24b, from which the corresponding values of $s_{2D}$ were taken. This last option gives a range and distribution of errors closely matching those
reported by B24 for the {\it Gaia} sample they use, which we have compared to.

Thus, the above procedure produces mock 'observational' wide binary samples restricted to the low acceleration $2<s_{2D}/kau<20$ range, incorporating the best
estimate of reality for this separation range: a statistical distribution of ellipticities from the direct empirical study of Hwang et al. (2022), a hidden tertiary
fraction consistent with the upper limits found in direct empirical studies of this quantity, e.g. Moe \& di Stefano (2017), Offner et al. (2023) (which have to be
understood as strict upper limits on $f_{\rm{multi}}$, as in that study it is the total multiplicity fraction that is inferred, and many such systems will be resolved
inner binaries removed from a wide binary test), an AQUAL underlying model as inferred independently in H22, H23, H34, C23, C24a and C24b, and crucially, the inclusion
of observational errors on the final $\tilde{v}$ sample.

The resulting 'observational' samples, which vary only due to the particular statistical realisation through which observational errors are added to a fixed AQUAL reality,
are then binned using exactly the same cell pixelation scheme of B24: 540 cells of size and shape taken from B24. An example of one such resulting pixelation scheme
is presented in the left panel of Fig. (6), which can be compared to the corresponding figure in B24. Then, we construct also two models matching the best-fit Newtonian
and MOND models reported in B24. The Newtonian model has a statistical ellipticity distribution with $\alpha=1.86$ and a hidden tertiary fraction of 0.699. The MOND
model assumes a statistical ellipticity distribution with $\alpha=1.96$ and a hidden tertiary fraction of 0.657. For both of these models, both of these parameters
are significantly larger than what is inferred directly by independent studies, as already mentioned, Hwang et al. (2022) constrains $\alpha<1.5$ and direct studies limit
the hidden tertiary fraction to below 0.5, including resolved inner binaries e.g. Moe \& di Stefano (2017), Offner et al. (2023). The parameters for the Newtonian and
MOND models described above are taken from Table 2 in B24, for their best-fit Newtonian and MOND models resulting from fitting the observed {\it Gaia} data to various
noise-free models.

Then, each of this two noise-free models is compared to 50 realisations of the 'observed' mock sample described, at each turn calculating ln P, the log likelihood
goodness-of-fit parameter used by B24 and described in eqs.(1), (2). For each assumed fixed observational Gaussian error amplitude, $(\sigma_{e}=0.04, \sigma_{e}=0.05,
\sigma_{e}=0.06, \sigma_{e}=0.07$ and $\sigma_{e}=0.08)$ and for the individual $\sigma_{ei}$ taken from the C24b sample described above, 50 statistical error realisations
were constructed of the underlying AQUAL model, which were then compared to both the noise-free Newtonian and the noise-free MOND models. Thus, for each assumed fixed
$\sigma_{e}$ value and the actual {\it Gaia} data error sample, 50 $ln P_{N}$ values for the Newtonian models, and 50 $ln P_{M}$ values for the MOND model were obtained.
The difference between them, $ln P_{N}-ln P_{M}$, which is the statistical comparison used in B24 to asses the preference of a Newtonian model over a MOND one,
was then calculated, and the mean value and the dispersion for each observational error assumption calculated. This final Newtonian vs. MOND values are displayed
in the right panel of Fig. (5).

The constant Gaussian error assumed is given in the x-axis, while the y-axis gives the mean and the 1$\sigma$ dispersion in
$ln P_{N}-ln P_{M}$ values resulting from comparing both the Newtonian and MOND noise-free models to the same AQUAL 'observational'
sample, after 50 different random observational error additions are performed on the AQUAL sample, blue dots with error bars. The horizontal
solid line at $(ln P_{N}-ln P_{M})=0$ divides cases with a preference for a Newtonian model, above this line, and those with a
preference for the MOND option, below the line. The dotted horizontal line gives $(ln P_{N}-ln P_{M})=175$, the value found in  B24
for the unique comparison of the {\it Gaia} data to their best-fit Newtonian and MOND noise-free models, having parameters as used here.
We see that as the errors increase, a slight bias appears such that the MOND model is increasingly less preferred over the Newtonian one,
for the reason described in connection to the PS23 study given in Section 3. In going to much larger values of $\sigma_{e}$ approaching
the value of 0.14 present in the data used by PS23, an AQUAL reality to which errors have been added becomes a better fit to a Newtonian
noise-free model than to a MOND noise-free one.

Beyond the actual values of the mean $(ln P_{N}-ln P_{M})$ comparisons shown, the most salient feature is the large 1$\sigma$ dispersion
obtained. As could have been anticipated from the very low mean cell occupancy numbers of 16 observed points per cell in the 540 cell
$(s_{2D}, \tilde{v})$ pixelation scheme in B24, particular error addition simulations result in very different prefered models
through the $(ln P_{N}-ln P_{M})$ goodness-of-fit parameter used by B24. Indeed, all experiments performed having constant Gaussian
errors above $\sigma_{e}=0.04$ result in 1$\sigma_{e}$  $(ln P_{N}-ln P_{M})$ distributions which encompass the $(ln P_{N}-ln P_{M})=175$
preference of the Newtonian model over the MOND one reported by B24, for the intrinsic AQUAL mock observational samples tested here
against noise-free models through the methods of B24.

Finally, for observationally selected Gaussian errors as derived from the {\it Gaia} sample, results are shown by the red dot with
error bar. The x-coordinate of this point was chosen arbitrarily so that the point would lie along a linear fit to the two constant amplitude error points flanking it.
Note that in this case a significant fraction of $\tilde{v}$ values have errors above the mean values, as shown in
Fig. 4 in B24. The actual error structure of the sample in this case, matching the one used by B24, includes many cases were the
errors are larger than the mean errors of the entire sample, and it is these points which primarily lead to the very large variance
in the recovered inferences. For this final case, the value of $(ln P_{N}-ln P_{M})=175$ reported by B24 is actually within 0.75$\sigma$
of the mean recovered  $(ln P_{N}-ln P_{M})$ value, when comparing an AQUAL reality to which observational errors have been added, to
the two noise-free models tested. Notice also that if one derives formal statistical significance values as $\sqrt{2 \Delta ln P}$,
as done in B24, the value of $(ln P_{N}-ln P_{M})=175$ reported by B24 corresponds to an 18.7 $\sigma$ preference for the Newtonian
noise-free model over the MOND noise-free one, when compared to the actual noisy reality of the {\it Gaia} data. The error bar for
the red point in the figure shows that within the dispersion, values of this statistical significance ranging from $(ln P_{N}-ln P_{M})=-552$, a 33.3 $\sigma$
preference for the MOND model over the Newtonian one, to $(ln P_{N}-ln P_{M})=280$, a 23.7 $\sigma$ preference for the Newtonian model over the MOND one,
are expected to occur.

Thus, the 18.7 $\sigma$ preference for the Newtonian noise-free model over the MOND noise-free one, when compared to the actual noisy
reality of the {\it Gaia} data reported in B24 is completely within expectations within an AQUAL reality, given the substantial variance of the problem, which the above
authors ignored. Formal statistical significance values are not relevant when a basic element of reality is absent from the models being tested
against real data, in this case observational noise. Notice that for the experiments presented in this sub-section, in the interest of a more complete coverage of the
($s_{2D}, \tilde{v}$) plane, we have started from a mock observational sample having 9,155 data points. Hence, in the case of B24 who use a slightly smaller unique
observational template of 8,611, all effects described in this sub-section will be slightly enhanced.

Note that although B24 report a 19 $\sigma$ preference for their best-fit noise-free Newtonian model over their best-fit noise-free MOND option, they also report a
16 $\sigma$ exclusion of any MOND alternative, and report this final number as their final significance level. As shown above, the variance due to the presence of noise
is sufficient to encompass the reported 19 $\sigma$ difference. As the pixelation scheme and the data template remain constant in B24, this will be even more so with
regards to the 16 $\sigma $ claimed exclusion. In the noise-variance dominated results which a rigorous accounting of the noise present imply, the MOND model will be
well within 1$\sigma$ of the variance.

\section{Conclusions}

We have summarised the results of recent papers by our two groups, H22, H23, H24, C23, C24a and C24b. These studies cover a range of various wide binary
sample selection strategies all extending from projected separations of 0.2-20 kau. Within the 0.2-2 kau high acceleration region MOND predictions
converge to Newtonian expectations. Indeed, without the use of any free parameters, our two groups, through the use of several distinct and independent
statistical approaches without including any parameter fitting, find complete accordance of observational inferences and Newtonian gravity. For the
low acceleration 2-20 kau region, again full accord is found in all of the above studies, but finding observations of {\it Gaia} wide binaries to be consistent
with MOND predictions, for elliptical orbits within a 1.5 boost to the gravitational constant, $G \to (1.5\pm 0.2) G$. In these studies, full accounting of the effects of
observational noise were included, as well as the elimination or careful modeling and calibration of hidden tertiaries and all projection effects of the problem.

On the other hand, PS23/B24 perform a simultaneous statistical fitting of 7 (3 in PS23) parameters describing the prevalence and orbital details of hidden
tertiaries, the orbital details of the wide binaries, the ellipticity distributions of both wide and tight orbits, the prevalence of line-of-sight
contaminants and the effective model of gravity, using samples limited to the $2000$ (5000 in PS23) $<(s_{2D}/au)<30,000$ (20,000 in PS23) range,
through a count-in-cells data to model comparison. It is important to note that in both PS23 and B24 the existence of observational errors is neglected
from the statistical fitting procedure. Also, no deep Newtonian region is included to calibrate any of the parameters being fitted. As these studies do
not cover the regime change reported by our groups, they are insensitive to any $s_{2D}$ dependence of the gravity law relevant to the wide binaries studied,
if occurring outside of the separation ranges covered.

In the cases of parameters for which previous and/or direct observational determinations exist, the best fit values found by B24 are
inconsistent, finding excessively large hidden tertiary fractions, excessively large $\alpha$ ellipticity parameters and excessively small numbers of
line-of-sight contaminants. As we have shown explicitly, working within excessively large hidden binary fractions will necessarily result in reduced
effective gravity inferences. The need for the presence of a deep Newtonian regime within the test performed, as a consistency check against this occurrence,
hence becomes clear.

As a consequence of working within the presence of an un-acknowledged observational noise component, the fundamental physical
scalings of the problem are absent from their results, with the inferred effective gravity parameter being uncorrelated with any of the three parameters
for which observational or previous determinations exist mentioned above. Forcing models which include no observational noise component to comply with data
that do, implies ignoring the presence of a substantial variance in the inferred results, given the very low mean cell occupancy numbers of only 16
present in B24. We show explicitly that for sample sizes and observational errors as present in B24, and using the exact same counts-in-cells strategy of those
authors to obtain log likelihood comparisons between data and models, the reported 19 sigma preference of their best-fit noise-free Newtonian model over
their noise-free MOND model, when both are compared to the {\it Gaia} data, which necessarily includes the presence of observational noise, lies well within
the dispersion expected. This dispersion ranges from a 33.3 $\sigma$ preference of the MOND model over the Newtonian one, to a 23.7 $\sigma$ preference of the Newtonian
one over the MOND one, when both are compared to an underling MOND reality including observational noise. The exclusion of noise from the models to be compared
to the data, as also happened in PS23, renders formal statistical significance values a poor estimate of the underlying physics of the problem.

\section*{acknowledgements}

The authors acknowledge the critical input of the scientific editor of this paper, and of three referees,
as important towards having reached a much more balanced, complete and hopefully useful, revised version of
our manuscript. Xavier Hernandez acknowledges helpful discussions with Luis Nasser on the topics treated in this paper.
This work has made use of data from the European Space Agency (ESA) mission {\it Gaia} ({https://www.
cosmos.esa.int/gaia}), processed by the {\it Gaia} Data Processing and Analysis Consortium (DPAC, {https://www.
cosmos.esa.int/web/gaia/dpac/consortium}). Funding for the DPAC has been provided by national institutions, in
particular the institutions participating in the {\it Gaia} Multilateral Agreement. Xavier Hernandez acknowledge
financial assistance from CONAHCYT and PAPIIT IN102624. This work was supported by the National Research Foundation
of Korea (grantNo. NRF-2022R1A2C1092306).

\section*{DATA AVAILABILITY}
All data used in this work will be shared on reasonable
request to the author.


\begin{thebibliography}{99}

  
\bibitem{} Banik I., Zhao H., 2018, MNRAS, 480, 2660

  

\bibitem{} Banik I., Pittordis C., Sutherland W., 2021, arXiv:2109.03827
  
\bibitem{} Banik I., Pittordis C., Sutherland W., Famaey B., Ibata R., Mieske S., Zhao H., 2024, MNRAS 527, 4573 

  

\bibitem{} Belokurov V., et al., 2020, MNRAS, 496, 1922

  
  

\bibitem{} Chae K.-H., 2023, ApJ, 952, 128

\bibitem{} Chae K.-H., 2024a, ApJ, 960, 114

\bibitem{} Chae K.-H., 2024b, ApJ, submitted (arXiv:2402.05720)
  

\bibitem{} Clarke, C. J. 2020, MNRAS, 491, L72


\bibitem{} El-Badry K., Rix H.-W., 2018, MNRAS, 480, 4884

\bibitem{} El-Badry, K., Rix, H.-W., Heintz, T. M. 2021, MNRAS, 506, 2269
  

\bibitem{} Hartman Z. D., Lepine S., Medan I. 2022, ApJ, 934, 72

\bibitem{} Hernandez X., Gilmore G., Valls-Gabaud, D., 2000, MNRAS, 317, 831
    
\bibitem{} Hernandez X., Jim\'{e}nez M. A., Allen C., 2012a, European Physical Journal C, 72, 1884

  
\bibitem{} Hernandez X., Cort\'{e}s R. A. M., Allen C. \& Scarpa R., 2019a, IJMPD, 28, 1950101

 

  

\bibitem{} Hernandez X., Cookson S., Cort\'{e}s R. A. M., 2022, MNRAS, 509, 2304 

\bibitem{} Hernandez X., 2023, MNRAS, 525, 1401  

\bibitem{} Hernandez X., Verteletskyi V., Nasser L., Aguayo-Ortiz A., 2024, MNRAS, 528, 4720

\bibitem{} Hwang H.-C., Ting Y.-S., Zakamska N. L., 2022, MNRAS, 512, 3383
    
\bibitem{} Jiang, Y. F., \& Tremaine, S., 2010, MNRAS, 401, 977
  

  


  
  

\bibitem{} Milgrom M., 1983, ApJ, 270, 365

\bibitem{} Milgrom M., 2010, MNRAS, 403, 886
  
\bibitem{} Moe, M., Di Stefano, R. 2017, ApJS, 230, 15

  

  



\bibitem{} Offner S. S. R., Moe M., Kratter K. M., Sadavoy S. I., Jensen E. L. N., Tobin J. J. 2023,  Protostars and Planets VII, ASP Conference Series, Vol. 534, p. 275 

\bibitem{} Penoyre Z., Belokurov V., Evans N. W., Everall A., Koposov S. E., 2020, MNRAS, 495, 321

\bibitem{} Pittordis C., Sutherland W., 2018, MNRAS, 480, 1778

\bibitem{} Pittordis C., Sutherland W., 2019, MNRAS, 488, 4740

\bibitem{} Pittordis C., Sutherland W., 2023, OJAp, 6, 4

\bibitem{} Press, W., Teukolsky, S., Vetterling, W., Flannery, B. 2007, Numerical Recipes
3rd Edition: The Art of Scientific Computing (Cambridge: Cambridge University Press)

  
\bibitem{} Shaya E. J., Olling R. P., 2011, ApJS, 192, 2
  
  

  


\bibitem{} {Tokovinin A. A., Smekhov M. G., 2002, A\&A, 382, 118}

\bibitem{} {Tokovinin A., Hartung M., Hayward T. L., 2010, AJ, 140, 510}
  
\bibitem{} {Tokovinin A. 2014, AJ, 147, 87}



  
\end{thebibliography}
\end{document}